\documentclass[conference]{IEEEtran}
\IEEEoverridecommandlockouts
\usepackage{comment}
\usepackage{cite}
\usepackage{amsmath,amssymb,amsfonts}
\usepackage{algpseudocode}
\usepackage[]{algorithm2e}
\usepackage{graphicx}
\usepackage{textcomp}
\usepackage{xcolor}
\usepackage{subcaption}
\usepackage{listings}
\usepackage{float}
\usepackage{balance}
\usepackage{booktabs}
\usepackage{hyperref}
\def\BibTeX{{\rm B\kern-.05em{\sc i\kern-.025em b}\kern-.08em
    T\kern-.1667em\lower.7ex\hbox{E}\kern-.125emX}}
\begin{document}

\title{Tuna: A Static Analysis Approach to Optimizing Deep Neural Networks}
\makeatletter
\newcommand{\linebreakand}{%
  \end{@IEEEauthorhalign}
  \hfill\mbox{}\par
  \mbox{}\hfill\begin{@IEEEauthorhalign}
}
\makeatother
\author{\IEEEauthorblockN{ Yao Wang}
\IEEEauthorblockA{\textit{Amazon Web Services} \\
wayao@amazon.com}
\and
\IEEEauthorblockN{ Xingyu Zhou}
\IEEEauthorblockA{\textit{Amazon Web Services} \\
zhoxingy@amazon.com}
\and
\IEEEauthorblockN{ Yanming Wang}
\IEEEauthorblockA{\textit{Amazon Web Services} \\
yanmwang@amazon.com}
\linebreakand
\IEEEauthorblockN{ Rui Li}
\IEEEauthorblockA{\textit{University of Utah} \\
lirui@cs.utah.edu}
\and
\IEEEauthorblockN{ Yong Wu}
\IEEEauthorblockA{\textit{Amazon Web Services} \\
yongwu@amazon.com}
\and
\IEEEauthorblockN{ Vin Sharma}
\IEEEauthorblockA{\textit{Amazon Web Services} \\
vinarm@amazon.com}
}

\maketitle

\begin{abstract}
We introduce Tuna, a static analysis approach to optimizing deep neural network programs. The optimization of tensor operations such as convolutions and matrix multiplications is the key to improving the performance of deep neural networks. Many deep learning model optimization mechanisms today use dynamic analysis, which relies on experimental execution on a target device to build a data-driven cost model of the program. The reliance on dynamic profiling not only requires access to target hardware at compilation time but also incurs significant cost in machine resources. We introduce an approach that profiles the program by constructing features based on the target hardware characteristics in order. We use static analysis of the relative performance of tensor operations to optimize the deep learning program. Experiments show that our approach can achieve up to $11\times$ performance compared to dynamic profiling based methods with the same compilation time. 
\end{abstract}

\section{Introduction}

Deep neural networks (DNN) are the mainstay of many applications including image classification, natural language processing, speech recognition, and automated driving. Cloud service providers are offering machine learning services that can compile, optimize, and deploy DNN on various target hardware. Optimizing a large set of DNN models, from convolutional neural networks to transformer-based networks, for a wide range of target hardware, from general purpose processors to AI accelerator ASICs, is constrained by two primary factors:\\ \\
\textbf{Compilation time and expense.} A long compilation time can be a bad user experience. In addition, the longer the compilation time, the greater the infrastructure costs.
\\ \\
\textbf{Cross compilation.} A compiler cannot assume that it has access to the target hardware and it must be able to cross-compile from a host machine with a different hardware than the target.
\\ \\
Current machine learning compilers use two common ways to generate high performance deep learning code for multiple target hardware. The first method is based on auto-tuning, as implemented in AutoTVM \cite{chen2018learning}, Ansor \cite{zheng2020ansor}, FlexTensor \cite{zheng2020flextensor}, and Halide auto-scheduler \cite{mullapudi2016automatically}. Systems using this approach usually conduct a machine-learning driven auto-tuning search across a large predefined program transformations space.  The system generates code samples and measures their performance on the target machine. It trains a machine learning model based on the collected measurement data, and use this model to guide subsequent code sample selection and code generation. The second method uses vendor-supplied kernel libraries ($\mathit{e.g.}$, TensorRT, OneDNN). This is widely adopted by current machine learning frameworks such as MXNet, PyTorch, and TensorFlow.  
Both these methods, however, have several drawbacks that restrict their application at scale. Although the auto-tuning approach significantly reduces the engineering effort needed to hand-craft efficient code for each target platform, training an effective cost model for that target hardware requires collecting profile data from the execution of entire DNN on a real device. This breaks the cross compilation constraints for production compilation service. Furthermore, auto-tuning based compilation requires large number of samplings in program transformation space to converge to local optima, which can be prohibitively long. For example, tuning TensorFlow SSD MobileNet for Amazon Graviton2 target with AutoTVM takes 240 hours. On the other hand, vendor kernel libraries provide efficient handcrafted kernel codes which do not require auto-tuning. However, vendor libraries don't cover every operation used in DNN models. Adding more kernels requires heavy engineering effort. In addition, one vendor library usually only supports a specific hardware architecture. We have to integrate various libraries into production service to support different target hardware, making it hard to manage service infrastructure and extend to new target.

In order to resolve the challenge of building a compilation service with limited compilation time and cross-compilation mechanism, this paper proposes a system, Tuna, that utilizes the static analysis and a combination of analytical cost modeling to optimizing DNN. Tuna has the following advantages comparing to auto-tuning based compilation:
\begin{itemize}
  \item Optimization is done with static analysis and no real hardware is required.
  \item Hardware-based cost model is transferable to different micro architectures. Only one single cost model is required for one major architecture, such as CPU and GPU.
  \item Unlike performance measurement which requires sequential execution on a target device, static analysis tasks can be fully paralleled on a multi-core CPU machine, which largely reduces total analyzing time.
\end{itemize}

Comparing to vendor kernel libraries, Tuna supports a large set of DNN kernels. Tuna provides a uniform compilation stack for various target hardware.

In this paper, we create cost model for both CPU and Nvidia GPU architectures. We evaluate our approach on cloud servers and edge devices. Experimental results show our approach largely reduces total compilation time while achieving better average performance comparing to dynamic profiled cost model.

\section{Overview}

Tuna is a deep learning kernel optimization framework fully relying on compile-time static analysis. Figure \ref{figs:overview} shows the overall architecture of Tuna. The system inputs are a tensor program $\mathit{e}$ together with corresponding loop transformation candidate space $\mathit{T_e}$. For a transformation $\mathit{t}$ $\in$ $\mathit{T_e}$, let $\mathit{i = g(e,t)}$ be the intermediate representation of transformed program. Code generator then generates low-level code $\mathit{a}$ ($\mathit{e.g.}$, assembly, PTX). Hardware-based program feature is extracted via $\mathit{pf = f(i,a)}$. Program performance score is calculated through $\mathit{c(pf)}$, with cost model function $\mathit{c}$. The objective of Tuna is formalized as the following: \begin{equation}
	\mathop{\arg\min}_{\mathit{t} \in \mathit{T_e}} \ \  \mathit{c(f(g(e,t),a))}.
\end{equation}

Tuna is built upon TVM \cite{chen2018tvm}, a deep learning compiler stack. While we reuse the tensor expression system, IR and code generator from TVM, the following two key components are the enhancements:\\
\\
\textbf{Hardware related cost model.} Existing search-based approaches mainly rely on high-level loop structure features, such as loop unrolling, loop vectorization and number of arithmetic operations. We argue that these high level loop features don't include hardware specifications which have great impact on kernel performance. Tuna analyzes both program IR and low-level generated codes to extract target hardware related features, such as number of SIMD instructions, CPU cache locality, GPU shared memory utilization, etc. With these low-level hardware features, Tuna cost model is accurate enough to predict the relative performance for a set of transformations of a tensor program, rather than relying on experimental execution on real hardware device.
\\ \\
\textbf{Multi-threaded search algorithm.} A multi-threaded search algorithm which can fully utilize multi-core CPU resources is the key to accelerate searching. We choose evolution strategies \cite{salimans2017evostrategy} as search algorithm. Evolution strategies is a parallel black-box optimization algorithm. The computation among each iteration of population can be executed simultaneously. With multi-core CPU machine or a multi-machine compilation system, the total searching time can be significantly reduced to be fit into the normal compilation flow.

\begin{figure}
    \centering
    \includegraphics[clip,angle=0, width=0.5\textwidth]{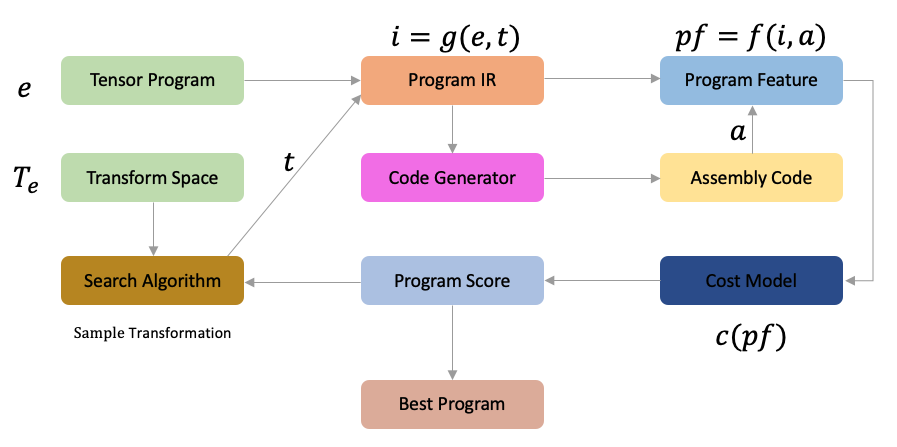}
    \caption{Overview of the static analysis approach to optimizing tensor programs.}
    \label{figs:overview}
\end{figure}

\section{Hardware Related Cost Model}
This section describes the analytical cost model to predict tensor program performance. Tuna system can be formulated as following: we extract a series of hardware related features, $\mathit{f_0, f_1, ..., f_n}$, which have great impact on deep learning tensor programs performance. These features can fall into two categories: \\\\
\textbf{Performance related instructions.} Tuna counts the number of low-level instructions that dominates tensor program performance, which mainly includes arithmetic and data movement instructions. We propose a unified method which jointly parses high-level program IR and low-level assembly code to get an accurate estimation of instruction numbers.\\\\
\textbf{General hardware features.} Hardware features such as cache locality and instruction level parallelism, have great impact on program performance. Specific algorithms are required to extract such features from program IR and assembly codes.\\\\
The program performance score is computed linearly with respect to features:
\begin{equation}
	\mathit{score = a_0 * f_0 + a_1 * f1 + ... + a_n * f_n}
\end{equation}
The coefficients $\mathit{a_0, a_1, ..., a_n}$ are generated for each hardware architecture through hardware instruction latency and empirical profiling data.

Tuna is generic enough to support different hardware architectures. This paper covers Intel CPU, ARM CPU and Nvidia GPU. We also investigate the transferability of the Tuna cost model across different micro architectures. Experiment results of high-end sever level chips versus resource-limited edge devices show that if two different micro architectures share the same set of Single-Instruction-Multiple-Data (SIMD) related instructions, a single cost model can be applied to both target platforms without modification.
\subsection{CPU cost model}
This section describes Tuna CPU cost model in detail. To achieve decent inference speed for deep learning program on modern CPU, it is common to do the following optimizations: thread-level parallelism to utilize multi-core CPU resources. Loop tiling to ensure good cache locality. Efficient utilization of SIMD registers to maximize vector computation throughput. Similarly, it's necessary to include these hardware features as parts of our CPU cost model to accurately analyze program performance. Tuna CPU cost model includes the following features:
\\ \\
\textbf{Number of SIMD instructions.} Compute-intensive tensor program performance is dominated by vector arithmetic and data movement instructions. For Intel AVX instruction set, $\mathit{vfmadd}$ and $\mathit{vmov}$ are the most common instructions in conv2d and dense operators, while for AARCH64 Neon $\mathit{fmla}$, $\mathit{ld}$ and $\mathit{st}$ are used. We parse the program IR and assembly codes to get the total number of these significant SIMD instructions.
\\ \\
\textbf{Estimation of L1 cache miss.} Cache locality is a key factor for tensor program on CPU. An analytical method is proposed in this paper to estimate L1 cache miss for a given program IR.
\\ \\
\textbf{Instruction level parallelism.} Comparing to handcraft kernel libraries which maximize fma instruction parallelism, search-based approach involves more complicated instruction arrangements. It is significant to evaluate whether the generated codes can keep CPU pipeline busy.

\subsubsection{Loop structure mapping}
We started from counting the number of SIMD instructions. Program IR represents the transformed tensor program in high level pattern, which preserves the complete loop structures. However, the actual SIMD instruction arrangement is opaque in high-level program IR, due to optimizations done in code generation process, such as register allocation and superword-level parallelism vectorization. On the other side, low-level assembly codes provide detailed instruction information in each basic block. However, it is difficult to restore original loop structure from assembly control-flow graph. It's infeasible to extract instruction information solely from either program IR or assembly code.

In this paper we propose an algorithm which jointly parse high-level program IR and low-level assembly code. The key idea is to extract full loop information from program IR and low-level instruction quantity from assembly code. A pattern matching is executed to match loop bodies with assembly basic blocks and calculate total number of SIMD instructions.

Algorithm \ref{algo:loopmap} shows the main procedure of jointly parsing. Loop blocks are extracted from program abstract syntax tree with pre-order depth-first-search. From assembly control flow graph, we identify a local basic block as a loop candidate with the following condition: Traversing from top to bottom of assembly code, there exists a jump instruction $\mathit{j}$ targeting a basic block $\mathit{LBB_x}$, and the position of $\mathit{LBB_x}$ is above $\mathit{j}$. Each pair of loop and basic block are then matched by checking whether they have the same iteration boundary. For matched basic blocks, we count the number of selected SIMD instructions. Finally, the algorithm calculates the total number of SIMD instructions for every loop block in original program IR.

\begin{algorithm}
\SetKwFunction{loopmap}{\textsc{Loop}-{Map}}
\SetKwFunction{fetchloop}{\textsc{Preorder}-{DFS}-{For}-{Loop}}
\SetKwFunction{idloopLBB}{\textsc{Identify}-{Loop}-{LBB}}
\SetKwFunction{matchloop}{\textsc{Pattern}-{Match}-{Loop}}
\SetKwFunction{countins}{\textsc{Count}-{Instruction}}
\Indm\loopmap{$IR, assembly$}
{\\
\Indp
  \emph{ForLoops} = \fetchloop{$IR$}\\
  \emph{LoopLBBs} = \idloopLBB{$assembly$}\\
  \emph{matchedIdx = 0}\\
  \emph{matchedLBBs} = []\\
  \emph{}
  \For{each basicBlock \textbf{in} LoopLBBs}{
     \emph{forLoop} = \emph{ForLoops[matchedIdx]}\\
     \If{\matchloop{forLoop, basicBlock}}{
         \emph{matchedLBBs.append(basicBlock)}\\
         \emph{matchedIdx} += 1
     }
  }
      \KwRet{\countins{$ForLoops, matchedLBBs$}}\\
  }

  \caption{Algorithm for mapping program IR and assembly code}
  \label{algo:loopmap}
\end{algorithm}
\subsubsection{Cache locality}

In this section, we developed an improved data locality model as an analyisis pass in TVM IR (TIR). This pass can analyze all kinds of computations supported by TVM in a short time. Therefore it provides a fast data locality approximation for the whole DNN network scheduling, and can be easily and systematically combined with other important models and passes together to guide the schedule selection of the whole network.  The main idea of the model is that we approximately estimate the data footprint and data movement volume required to move into the cache of the object code.  The object code was abstracted as a tree consists of loop-nodes and access-nodes.  All loop-nodes are non-leaf nodes and carry all information of a loop statement.  All access-nodes are leaf nodes that represent either a tensor load access or a tensor store access.  The data footprint and data movement are calculated by traversing the bottom to the top of the tree.
\begin{lstlisting}[language=C, float, basicstyle=\footnotesize, label={lst:2mmtile}, caption={Fused and tiled two matrix multiplication}, captionpos=b, belowskip=-1em]
// Ni/Nj/Nk  are perfect multiples of Ti/Tj/Tk 
for(it = 0; it < Ni; it+=Ti)
  for(jt = 0; jt < Nj; jt+=Tj)
    for(k = 0; k < Nk; k++)
    | for(i1 = 0; i1 < Ti; i1++)
    |   for(j1 = 0; j1 < Tj; j1++)
    |     C[i+it][j+jt]+=
    |     A[i+it][k]*B[k][j+jt];
    for(l = 0; l < Nl; l++)
      for(i2 = 0; i2 < Ti; i2++)
        for(j2 = 0; j2 < Tj; j2++)
          E[i+it][l]+=
          C[i+it][j+jt]*D[j+jt][l]
\end{lstlisting} 

We use Two Matrix Multiply (2MM) as an example to demonstrate the method for building data movement.  Listing \ref{lst:2mmtile} shows an example of fused and tiled 2MM code.  In this example the first Matmul uses $\mathit{i}$ and $\mathit{j}$ as free index and perform contraction over $\mathit{k}$.  The second Matmul  uses $\mathit{i}$ and $\mathit{l}$ as free index and perform contraction over $\mathit{j}$.  The tiling loop of $\mathit{i}$ and $\mathit{j}$ are tiled and fused together, and other loops are non-tiled and non-fused. The cache capacity and values of all tile-size and problem size are known before the analysis starts.

We assume the cache capacity $ \mathit{S}$ is enough to store all data footprints below tiling loops, and is not enough to store data footprints of any tiling loop.  This means $\mathit{Ni} > \mathit{S}$ and $\mathit{Nj} > \mathit{S}$, but $\mathit{S} > \mathit{Ti}\mathit{Tj} + \mathit{Ti}\mathit{Nl} + \mathit{Tj}\mathit{Nl}+ \mathit{Tj}\mathit{Nk} + \mathit{Ti}\mathit{Nk}$.

The data footprint and data movement are computed from leaf node to the root of the tree.  Since all leaf nodes are tensor accesses, the footprint and data movement for leaf nodes are both $1$.  The data footprint of a loop node is the number of distinct data elements accessed during all iterations of this loop.  The data movement of a loop node is calculated based on its footprint and cache capacity.  If the footprint of a single loop iteration is smaller than cache capacity, the data movement of this loop node is equal to its node footprint. Otherwise, the data movement of this loop node is evaluated as product of the number of iterations and the data movement of its single iteration.  The data movement of single iteration is correlated to data movement of sub nodes, which is computed earlier in the bottom-up procedure.

In the 2MM example, based on our capacity assumption, the footprint of a single iteration of loop $\mathit{jt}$ is  $\mathit{Ti}\mathit{Tj} + \mathit{Ti}\mathit{Nl} + \mathit{Tj}\mathit{Nl}+ \mathit{Tj}\mathit{Nk} + \mathit{Ti}\mathit{Nk} $ which fits in cache.  for all sub-loop nodes of $\mathit{jt}$, the data movement is equal to data footprint.  For loop $\mathit{jt}$, since its single iteration footprint fit in cache, when the control flow goes to the next iteration, tensor $\mathit{A}$ and $\mathit{E}$ could get reuse because their access functions do not include index $\mathit{jt}$. Therefore, the data movement of loop $\mathit{jt}$ is $\mathit{Ti}\mathit{Nj} + \mathit{Ti}\mathit{Nl} + \mathit{Nj}\mathit{Nl}+ \mathit{Nj}\mathit{Nk} + \mathit{Ti}\mathit{Nk}$, which is still equal to its footprint.  However, for loop $\mathit{it}$, even a single iteration footprint does not fit in cache.  So the data movement of loop $\mathit{it}$ will be the product of number of iterations and data movement of a single iteration, which is $( \mathit{Ti}\mathit{Nj} + \mathit{Ti}\mathit{Nl} + \mathit{Nj}\mathit{Nl}+ \mathit{Nj}\mathit{Nk} + \mathit{Ti}\mathit{Nk}) * \mathit{Ni}/\mathit{Ti}$.  Figure \ref{fig:2mm} shows the loop structure and data movement calculation of the 2MM example.

Algorithm \ref{algo:visitnode} shows the main procedure of visiting each node in the TIR node tree.  When visiting a loop node, the algorithm recursively visits all its children, and compute the union of data footprint of a single iteration.  The algorithm will detect whether the union of data footprint can fit in cache, and track the reuse status of each tensor.  If the data footprint of that iteration fits in cache, the data movement volume is same as its footprint.  Otherwise, the algorithm will use the tracked reuse status information to calculate the data movement.  If the reuse status of the tensor is true, the movement volume of that tensor will be equal to footprint, otherwise the movement volume will be the movement volume of a single iteration multiply the trip-count.  The reuse status will be true at the leaf node by default.  While our analysis move from the bottom to top, the reuse status will be flipped to false if the following conditions are true. The first case is if tensor footprint exceed cache.  The second case is if there exists a set of continuous loop nodes that do not access this tensor, such that their footprints exceed cache. Both implies the reuse distance of the tensor under discussion exceed the cache capacity.  The whole analysis module is implemented by using Integer Set Library \cite{verdoolaege2010isl}.  

\begin{algorithm}
\SetKwFunction{createSet}{\textsc{Create}-{IntegerSet}}
\SetKwFunction{expandDfp}{\textsc{Estimate}-{Dfp}}
\SetKwFunction{visitnode}{\textsc{Visit}-{Node}}
\SetKwFunction{updateReuse}{\textsc{Update}-{Reuse}-{Status}}
\Indm\visitnode{$node, cache$}
{\\
\Indp
  \If{node is AccessNode}{
  \label{algo:visitnode:leaf}
  \KwRet{Dmov=1, Dfp=1, DataSpace}} 
  \ElseIf{node is LoopNode}{
  \label{algo:visitnode:loop}
  \emph{subDspace, subDfp, subDmov} = []\\
   \For{each childNode}{
      \label{algo:visitnode:child}
      \emph{dspace}, \emph{dfp}, \emph{dmov} = \visitnode{childNode, cache}\\
      append \emph{dspace, dfp, dmov} to \emph{subDspace, subDfp, subDmov}
   }
   \emph{DataSpace} = \createSet{subDspace}\\
   \If{DataSpace.cardinality $>$ cache}{
     \label{algo:visitnode:exceed}
      \emph{Dmov} = 0\\
      \For{each tensor ts}{
         get \emph{ts.dmov, ts.dspace, ts.expr} from \emph{subDmov, subDspace}\\
         \If{ts.reuse is True}{
              \emph{Dmov} += \expandDfp{node.iterVar, ts.dspace}
              \updateReuse{ts}
        }
        \Else{
              \emph{Dmov} += \emph{ts.dmov} * \emph{node.tripcount}
        }

      }
      \KwRet{Dmov, Dfp=DataSpace.cardinality, DataSpace}
   }
   \Else{
      \label{algo:visitnode:notexceed}
     \KwRet{Dmov=Dfp, Dfp=DataSpace.cardinality, DataSpace}\\
   }
  }
  
}
  \caption{Algorithm for modeling data movement at each TIR tree node}
  \label{algo:visitnode}
\end{algorithm}

\begin{figure}
    \centering
    \includegraphics[clip,angle=0, width=0.5\textwidth]{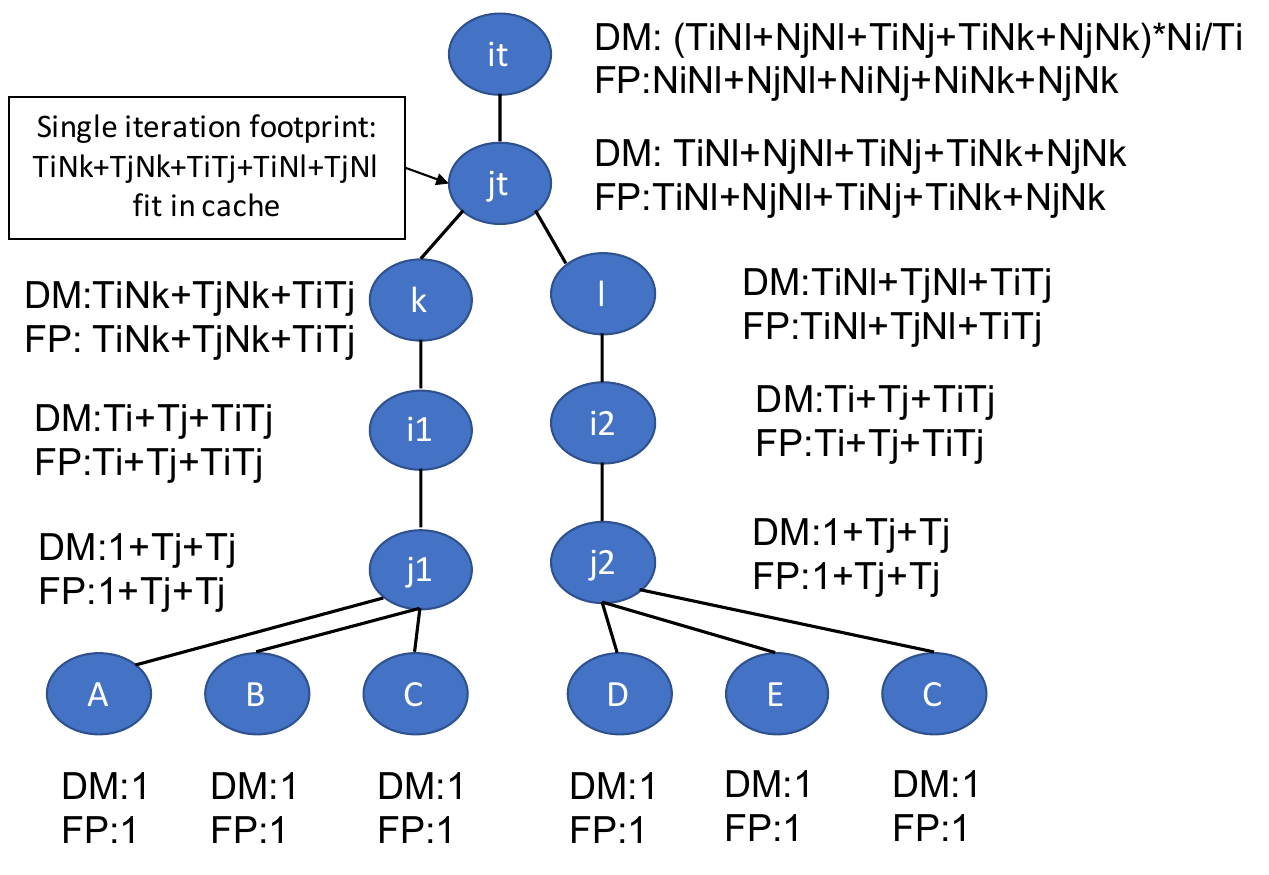}
    \caption{Loop structure and data movement of 2MM}
    \label{fig:2mm}
\end{figure}

\subsubsection{Instruction level parallelization}
Modern CPU utilizes pipelines and out-of-order execution to increase the instruction level parallelism (ILP) and hide the latency.  Therefore, to achieve high performance, the generated assembly code should allow the processor issue as much instructions as possible to keep the pipeline busy. Since the TVM framework includes more types of operators which will generates more types of instructions, a specific model for specific operator is not adequate to apply on the TVM-generated code.  To solve this problem, we propose a static analysis model to estimate the instruction level efficiency. 

The key idea of the model is to design a simplified fast out-of-order instruction scheduler that schedule instructions in each basic block.  The scheduler consists of two major components, the data dependency builder and the instruction scheduler.  The data dependency builder first scan the whole basic block, and creates two instruction dependency graph for true dependency and false dependency respectively.  Then the instruction scheduler will schedule the instructions based on the dependency graph and hardware specifications such as instruction latency and number of different processing unit.  During the scheduling, a timestamp will be assigned to each instruction, indicating the time point of the instruction starting executed. The first ready-to-execute instruction will be scheduled at cycle zero, and total cycles required for finalize all instructions will be used as the ILP cost of this basic block.  We calculate the product of ILP cost and number of executions for a single basic block, and add up all products of all basic block.  The summation is used as the ILP cost of the whole program.

During the scheduling, the scheduler will manage two different hazard, the structural hazard and data hazard.  The structure hazard is controlled by limiting the maximum number of instruction issued at each cycle.  If the maximum number of issued instruction reached the number of processing unit, the next instruction to be issued will be delayed to next cycle. The data hazard is identified by analyzing the dependency graph.  If there is a read-after-write (RAW) dependency between two instructions, the consumer instruction should be scheduled after the producer instruction finishing execution.  If there is a write-after-read (WAR) dependency or write-after-write dependency, the latter instruction who writes to the resource cannot be scheduled before the prior instruction.
\subsection{Nvidia GPU cost model}
This section describes our Nvidia GPU cost model in detail. To achieve efficient execution, GPU tensor program needs to exhibit decent thread-level parallelism to utilize GPU thread resources. Data locality is also significant for shared memory efficiency. Our GPU cost model includes the following features:
\\ \\
\textbf{Number of PTX instructions}.  Similar to CPU assembly codes, we select $\mathit{fma}$, $\mathit{ld}$ and $\mathit{st}$ as the most significant instructions for tensor programs. We parse the Nvidia GPU PTX code to get the total number of these instructions.
\\ \\
\textbf{Thread level parallelism}. Utilization of GPU threads largely determines the performance of GPU kernels. We evaluate several aspects that directly affect GPU thread-level parallelism: number of operations in a single thread, Streaming Multiprocessor (SM) occupancy, warp latency hiding and shared memory bank conflict.
\subsubsection{Loop structure mapping}
The NVCC compilation process unrolls small loops with a known trip count by default, which makes it hard to identify the corresponding loop structure in high-level program IR from the low-level PTX code. 

In this paper we propose an algorithm that parses the PTX code, identifies loop structures and calculates total number of instructions. The key idea is to identify the loop structure from PTX code and loop iterations from registers.

Algorithm 3 shows the key idea of how we get the number of instructions. We adopt the same idea from Algorithm 1 of identifying loop structures in assembly code since PTX code has similar condition and jump structures. After we have the loop structures, we maintain a register initial value map and register value update map by parsing the PTX code. Since we already know the loop structure, once we reach the line with eligible condition check, we know it is for a certain loop. Based on the condition check, we get the register used for comparing and the end condition. From the two maps we maintain, we can easily calculate the loop iterations by initial value, update value and end condition. Finally, the algorithm calculates the total number of PTX instructions for every loop.

\begin{algorithm}
\SetKwFunction{loopmapptx}{\textsc{Loop}-{Map}-{PTX}}
\SetKwFunction{idloopBB}{\textsc{Identify}-{Loop}-{BB}}
\SetKwFunction{registerextract}{\textsc{Register}-{Match}-{Loop}}
\SetKwFunction{countins}{\textsc{Count}-{Instruction}}
\SetKwFunction{calculate}{\textsc{Get}-{Iterations}}
\Indm\loopmapptx{$PTX$}
{\\
\Indp
  \emph{LoopBBs} = \idloopBB{$PTX$}\\
  \emph{regInitMap, regUpdateMap} = \registerextract{}\\
  \emph{LoopIterations} = []\\
  \For{each basicBlock \textbf{in} LoopBBs}{
    \emph{loopIteration} = \calculate{$regInitMap, regUpdateMap$}\\
    \emph{LoopIterations.append(loopIteration)}\\
  }
      \KwRet{\countins{$LoopBBs, LoopIterations$}}\\
  }

  \caption{Algorithm for identifying loop iterations in PTX code}
  \label{algo:loopmapptx}
\end{algorithm}
\subsubsection{Thread level parallelization}

In this section we discuss the details about features regarding to thread-level parallelism.
\\ \\ 
\textbf{Workload per thread} We count the total number of PTX instructions for a single thread with Algorithm \ref{algo:loopmapptx}. We calculate the total number of cycles based on instruction cycles from \cite{arafa2019ppt}:\\\\
\begin{equation}
\sum_{i=1}^{PTX Instruction} Count(i)*Cost(i)\
\end{equation}
The workload per thread is represented by the total number of cycles.
\\ \\
\textbf{Streaming Multiprocessor (SM) occupancy} It is important to keep all SMs busy for compute-intensive tensor programs. We check the total number of thread blocks to determine whether it is greater than the total number of SMs, so that all SMs have at least one block to execute. A penalty is added if the total number of thread blocks is too small.
\\ \\
\textbf{Warp latency hiding} Warp scheduling is a vital mechanism to hide GPU instruction latency. With more warps on each SM, GPU warp scheduler has a better chance to hide memory latency, thus achieving better performance. We calculate the maximum number of thread blocks that can be concurrently scheduled in one SM by checking the number of registers and shared memory usage per block. These information can be extracted with \emph{nvcc ptxas-option} command.
\\ \\
\textbf{Shared memory bank conflict}
For modern Nvidia GPUs (compute capability $>=$ 5.0), shared memory has 32 banks that are organized such that successive 32-bit words map to successive banks. Each bank has a bandwidth of 32 bits per clock cycle and any shared memory load or store of n addresses that spans $n$ distinct memory banks can be served simultaneously, yielding an effective bandwidth that is $n$ times as high as the bandwidth of a single bank. However, if multiple addresses of a memory request map to the same memory bank, the accesses are serialized. The hardware splits a memory request that has bank conflicts into as many separate conflict-free requests as necessary, decreasing the effective bandwidth by a factor equal to the number of separate memory requests. The one exception here is when multiple threads in a warp address the same shared memory location, resulting in a broadcast. In this case, multiple broadcasts from different banks are coalesced into a single cast from the requested shared memory locations to the threads. To incorporate the effect of bank conflict, we first numerically evaluate the shared memory access indices of all threads in the first warp from the IR to compute the actual shared memory throughput. We then use the ratio between actual shared memory throughput and requested shared memory throughput to adjust the number of shared memory operations.
\section{Search algorithm}

\begin{table*}[htp]
\begin{subtable}{\textwidth}
\centering
	\resizebox{\textwidth}{!}{%
        \begin{tabular}{ccccccccc}
        \toprule
        Unit: ms & TF SSD MobileNet & TF SSD Inception & PT ResNet50 & PT Bert \\ \hline \\[-1.75ex]
        Framework & 22.09 & 24.29 & 31.26 & 253.04 \\
        AutoTVM Partial & 30.41 & 47.7 & 11.47 & 86.39 \\
        AutoTVM Full & 22.7 & 30.29 & 6.37 & 16.66 \\
        Tuna & 23.3 & 30.16 & 6.85 & 15.12 & \\ \hline
        \end{tabular}
    }
\caption{Entire network performance on a system with Intel Xeon Platinum 8124M CPU}
\label{tbl:intel}
\end{subtable}\\\\

\begin{subtable}{\textwidth}
\centering
	\resizebox{\textwidth}{!}{%
        \begin{tabular}{ccccccccc}
        \toprule
        Unit: ms & TF SSD MobileNet & TF SSD Inception & PT ResNet50 & PT Bert \\ \hline \\[-1.75ex]
        Framework & 37.15 & 43.39 & 307.55 & 56.7 \\
        AutoTVM Partial & 46.95 & 75.37 & 195.38 & 63.62 \\
        AutoTVM Full & 29.55 & 45.56 & 16.11 & 15.11 \\
        Tuna & 30.24 & 44.3 & 17.77 & 16.13 & \\ \hline
        \end{tabular}
    }
\caption{Entire network performance on a system with AWS Graviton2 ARM CPU}
\label{tbl:amd}
\end{subtable}\\\\

\begin{subtable}{\textwidth}
\centering
	\resizebox{\textwidth}{!}{%
        \begin{tabular}{ccccccccc}
        \toprule
        Unit: ms & TF SSD MobileNet & TF SSD Inception & PT ResNet50 & PT Bert \\ \hline \\[-1.75ex]
        AutoTVM Partial & 3499.72 & 4487.97 & 7111.88 & 1058.85 \\
        AutoTVM Full & 618.45 & 1419.83 & 1114.69 & 574.37 \\
        Tuna & 614.43 & 1389.35 & 1259.88 & 541.17 & \\ \hline
        \end{tabular}
    }
\caption{Entire network performance on a system with ARM Quad-core Cortex-A53 64-bit CPU(Acer aiSage)}
\label{tbl:amd}
\end{subtable}\\\\

\begin{subtable}{\textwidth}
\centering
	\resizebox{\textwidth}{!}{%
        \begin{tabular}{ccccccccc}
        \toprule
        Unit: ms & TF SSD MobileNet & TF SSD Inception & PT ResNet50 & PT Bert \\ \hline \\[-1.75ex]
        Framework & 31.65 & 33.2 & 8.65 & 16.34 \\
        AutoTVM Partial & 44.77 & 70.97 & 105.16 & 13.03 \\
        AutoTVM Full & 27.43 & 30.69 & 2.34 & 3.24 \\
        Tuna & 28.45 & 34.87 & 2.65 & 4.62 & \\ \hline
        \end{tabular}
    }
\caption{Entire network performance on a system with Nvidia V100 GPU}
\label{tbl:amd}
\end{subtable}\\\\

\begin{subtable}{\textwidth}
\centering
	\resizebox{\textwidth}{!}{%
        \begin{tabular}{ccccccccc}
        \toprule
        Unit: ms & TF SSD MobileNet & TF SSD Inception & PT ResNet50 & PT Bert \\ \hline \\[-1.75ex]
        Framework & 126.7 & 124.13 & 26.04 & 19.66 \\
        AutoTVM Partial & 202.21 & 426.44 & 21.16 & 51.32 \\
        AutoTVM Full & 50.78 & 57.25 & 15.21 & 7.64 \\
        Tuna & 59.81 & 80.05 & 18.82 & 11.49 & \\ \hline
        \end{tabular}
    }
\caption{Entire network performance on a system with Nvidia Jetson AGX Xavier GPU}
\label{tbl:amd}
\end{subtable}

\caption{Entire network performance of \emph{Tuna} and the selected baselines.}
\label{tbl:network}
\end{table*}

\begin{table*}[htp]
\begin{subtable}{\textwidth}
\centering
	\resizebox{\textwidth}{!}{%
        \begin{tabular}{ccccccccc}
        \toprule
        Unit: hour & TF SSD MobileNet & TF SSD Inception & PT ResNet50 & PT Bert \\ \hline \\[-1.75ex]
        AutoTVM & 53 & 50 & 12 & 2.92 \\
        Tuna & 0.7 & 1 & 0.13 & 0.012 & \\ \hline
        \end{tabular}
    }
\caption{Entire network compilation time for Intel Xeon Platinum 8124M CPU.}
\label{tbl:intel}
\end{subtable}\\\\

\begin{subtable}{\textwidth}
\centering
	\resizebox{\textwidth}{!}{%
        \begin{tabular}{ccccccccc}
        \toprule
        Unit: hour & TF SSD MobileNet & TF SSD Inception & PT ResNet50 & PT Bert \\ \hline \\[-1.75ex]
        AutoTVM & 240 & 280 & 49 & 3.2 \\
        Tuna & 3 & 5 & 0.45 & 0.062 & \\ \hline
        \end{tabular}
    }
\caption{Entire network compilation time for AWS Graviton2 ARM CPU.}
\label{tbl:amd}
\end{subtable}\\\\

\begin{subtable}{\textwidth}
\centering
	\resizebox{\textwidth}{!}{%
        \begin{tabular}{ccccccccc}
        \toprule
        Unit: hour & TF SSD MobileNet & TF SSD Inception & PT ResNet50 & PT Bert \\ \hline \\[-1.75ex]
        AutoTVM & 299 & 316 & 64 & 5.14 \\
        Tuna & 3 & 5 & 0.45 & 0.062 & \\ \hline
        \end{tabular}
    }
\caption{Entire network compilation time for ARM Quad-core Cortex-A53 64-bit CPU(Acer aiSage).}
\label{tbl:amd}
\end{subtable}\\\\

\begin{subtable}{\textwidth}
\centering
	\resizebox{\textwidth}{!}{%
        \begin{tabular}{ccccccccc}
        \toprule
        Unit: hour & TF SSD MobileNet & TF SSD Inception & PT ResNet50 & PT Bert \\ \hline \\[-1.75ex]
        AutoTVM & 93.5 & 167.5 & 37 & 4.4 \\
        Tuna & 1.6 & 1.04 & 0.13 & 0.05 & \\ \hline
        \end{tabular}
    }
\caption{Entire network compilation time for AWS P3 V100 GPU.}
\label{tbl:amd}
\end{subtable}\\\\

\begin{subtable}{\textwidth}
\centering
	\resizebox{\textwidth}{!}{%
        \begin{tabular}{ccccccccc}
        \toprule
        Unit: hour & TF SSD MobileNet & TF SSD Inception & PT ResNet50 & PT Bert \\ \hline \\[-1.75ex]
        AutoTVM & 202.5 & 356.3 & 83 & 4.4 \\
        Tuna & 1.7 & 1.05 & 0.13 & 0.05 & \\ \hline
        \end{tabular}
    }
\caption{Entire network compilation time for Nvidia Jetson AGX Xavier GPU.}
\label{tbl:amd}
\end{subtable}

\caption{Entire network compilation time of \emph{Tuna} VS \emph{AutoTVM}.}
\label{tbl:time}
\end{table*}

\begin{table*}[htp]
\begin{subtable}{\textwidth}
\centering
	\resizebox{\textwidth}{!}{%
        \begin{tabular}{ccccccccc}
        \toprule
        Unit: dollar & TF SSD MobileNet & TF SSD Inception & PT ResNet50 & PT Bert \\ \hline \\[-1.75ex]
        AutoTVM & 81.09 & 76.5 & 18.36 & 4.47 \\
        Tuna & 2.86 & 4.08 & 0.53 & 0.05 & \\ \hline
        \end{tabular}
    }
\caption{Entire network compilation cost for Amazon EC2 C5.9xlarge(price \$1.53 per hour).}
\label{tbl:intel}
\end{subtable}\\\\

\begin{subtable}{\textwidth}
\centering
	\resizebox{\textwidth}{!}{%
        \begin{tabular}{ccccccccc}
        \toprule
        Unit: dollar & TF SSD MobileNet & TF SSD Inception & PT ResNet50 & PT Bert \\ \hline \\[-1.75ex]
        AutoTVM & 147.84 & 172.48 & 30.18 & 1.97 \\
        Tuna & 12.24 & 20.4 & 1.84 & 0.25 & \\ \hline
        \end{tabular}
    }
\caption{Entire network compilation cost for AWS Graviton2 ARM CPU(price \$0.616 per hour).}
\label{tbl:amd}
\end{subtable}\\\\

\begin{subtable}{\textwidth}
\centering
	\resizebox{\textwidth}{!}{%
        \begin{tabular}{ccccccccc}
        \toprule
        Unit: dollar & TF SSD MobileNet & TF SSD Inception & PT ResNet50 & PT Bert \\ \hline \\[-1.75ex]
        AutoTVM & 286.32 & 512.61 & 113.22 & 13.46 \\
        Tuna & 6.53 & 4.24 & 0.53 & 0.2 & \\ \hline
        \end{tabular}
    }
\caption{Entire network compilation cost for Amazon EC2 P3.2xlarge(price \$3.06 per hour).}
\label{tbl:amd}
\end{subtable}

\caption{Entire network compilation cost of \emph{Tuna} VS \emph{AutoTVM}.}
\label{tbl:cost}
\end{table*}

\begin{figure}[!htb]
	\centering
	\begin{subfigure}{0.47\textwidth}
	\begin{minipage}[t]{\textwidth}
	    \centering
		\includegraphics[width=\linewidth]{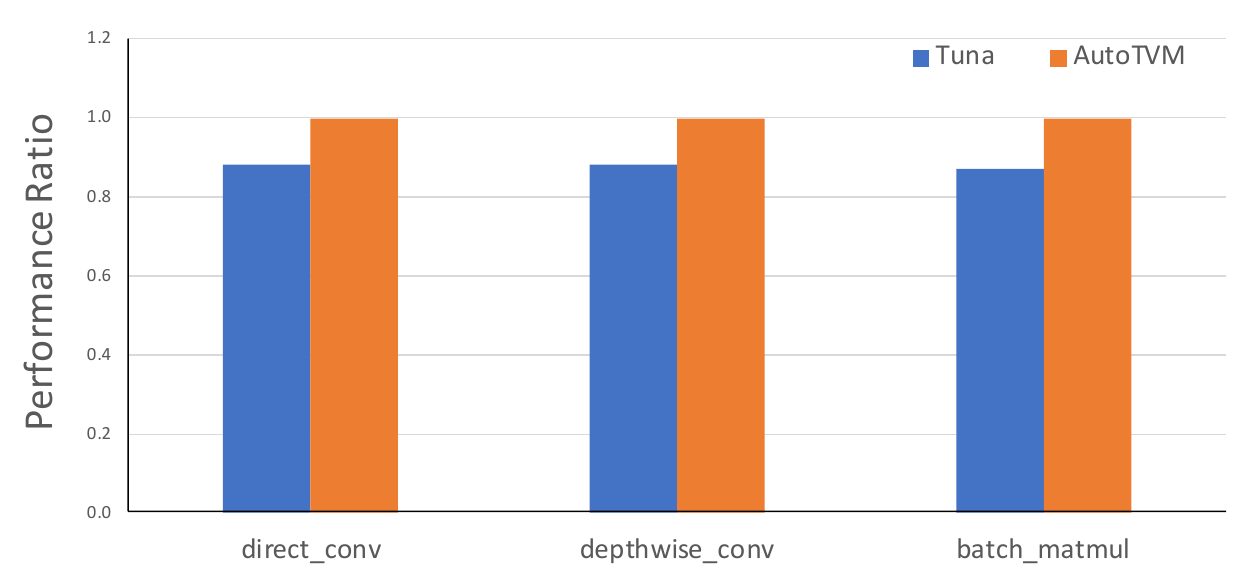}
		\caption{Top-10 Performance Ratio on Intel Xeon Platinum 8124M CPU}
		\label{fig:single:top10-intel}
	\end{minipage}%
	\end{subfigure}
	\\
	\begin{subfigure}{0.47\textwidth}
	\begin{minipage}[t]{\textwidth}
		\centering
		\includegraphics[width=\linewidth]{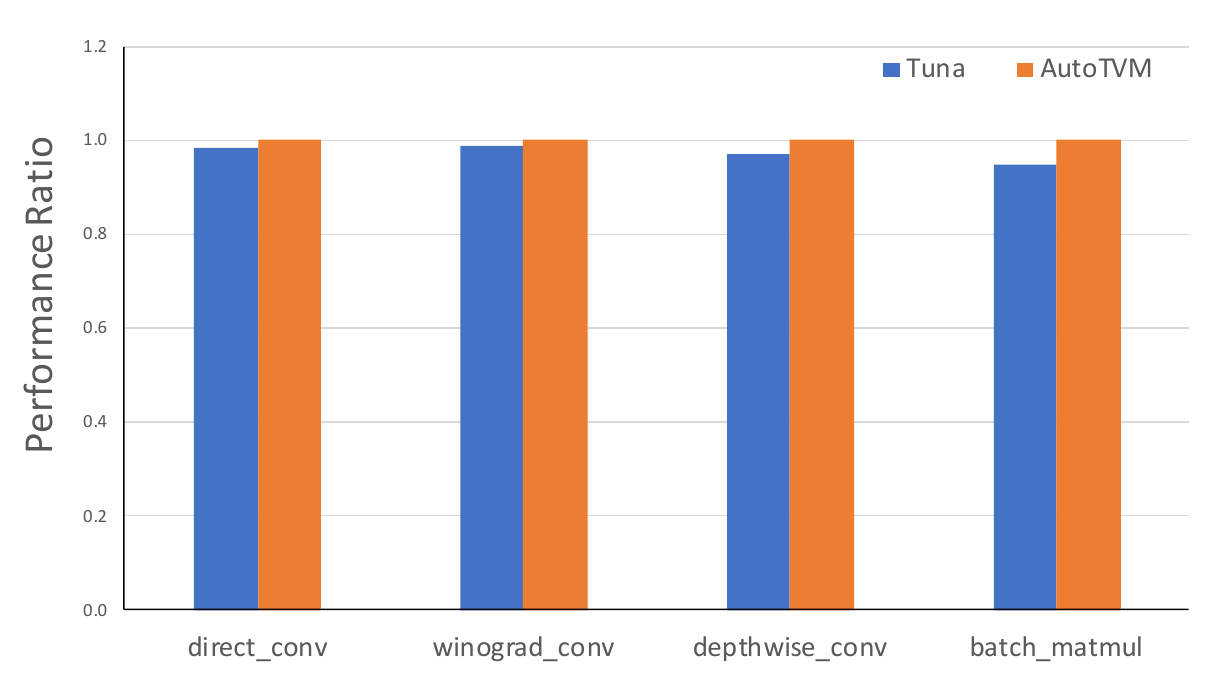}
		\caption{Top-10 Performance Ratio on AWS Graviton2 ARM CPU}
		\label{fig:single:top10-intel}
	\end{minipage}%
	\end{subfigure}
	\\
	\begin{subfigure}{0.47\textwidth}
	\begin{minipage}[t]{\textwidth}
		\centering
		\includegraphics[width=\linewidth]{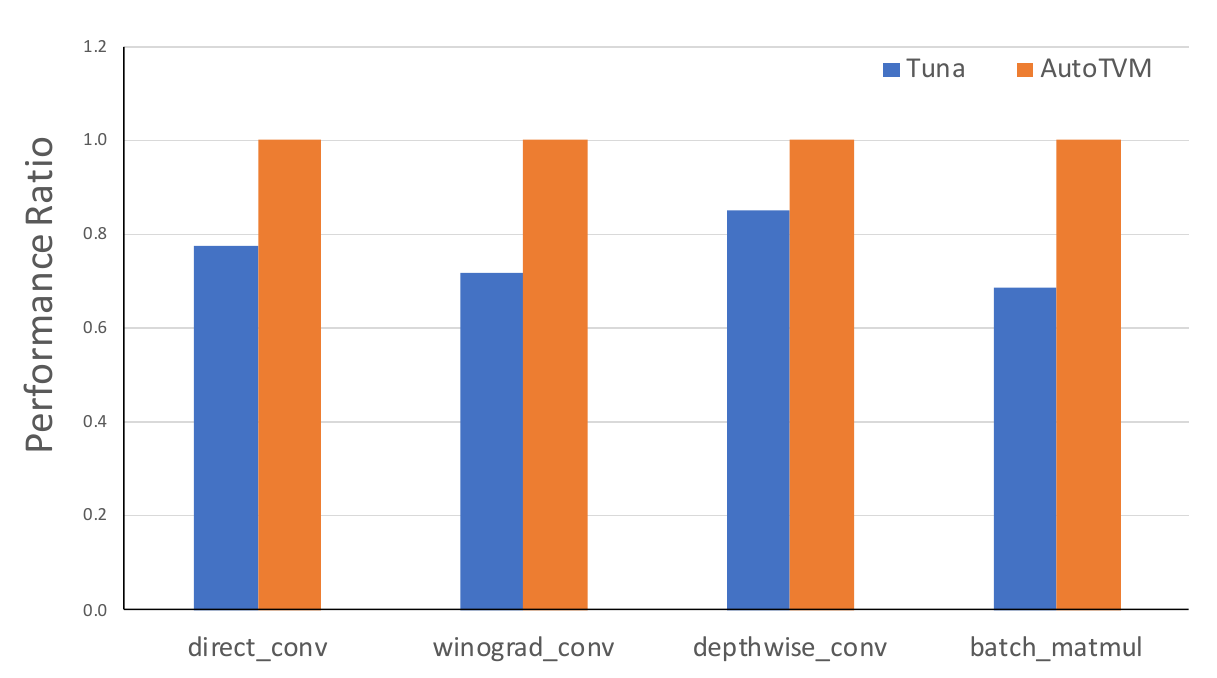}
		\caption{Top-10 Performance Ratio on AWS P3 V100 GPU}
		\label{fig:single:top10-intel}
	\end{minipage}%
	\end{subfigure}
	\caption{Top10 performance ratio for single operators from Tuna VS AutoTVM.}
	\vspace{-1em}
	\label{fig:single10}
\end{figure}

\begin{figure}[!htb]
	\centering
	\begin{subfigure}{0.47\textwidth}
	\begin{minipage}[t]{\textwidth}
		\centering
		\includegraphics[width=\linewidth]{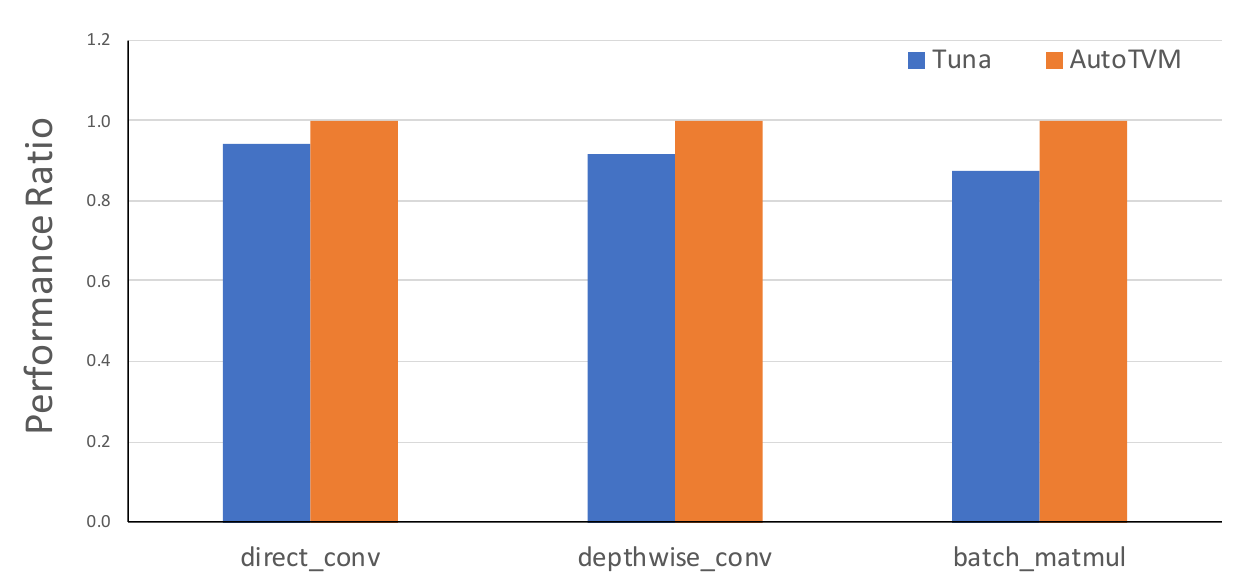}
		\caption{Top-50 Performance Ratio on Intel Xeon Platinum 8124M CPU}
		\label{fig:single:top50-intel}
	\end{minipage}%
	\end{subfigure}
	\\
	\begin{subfigure}{0.47\textwidth}
	\begin{minipage}[t]{\textwidth}
		\centering
		\includegraphics[width=\linewidth]{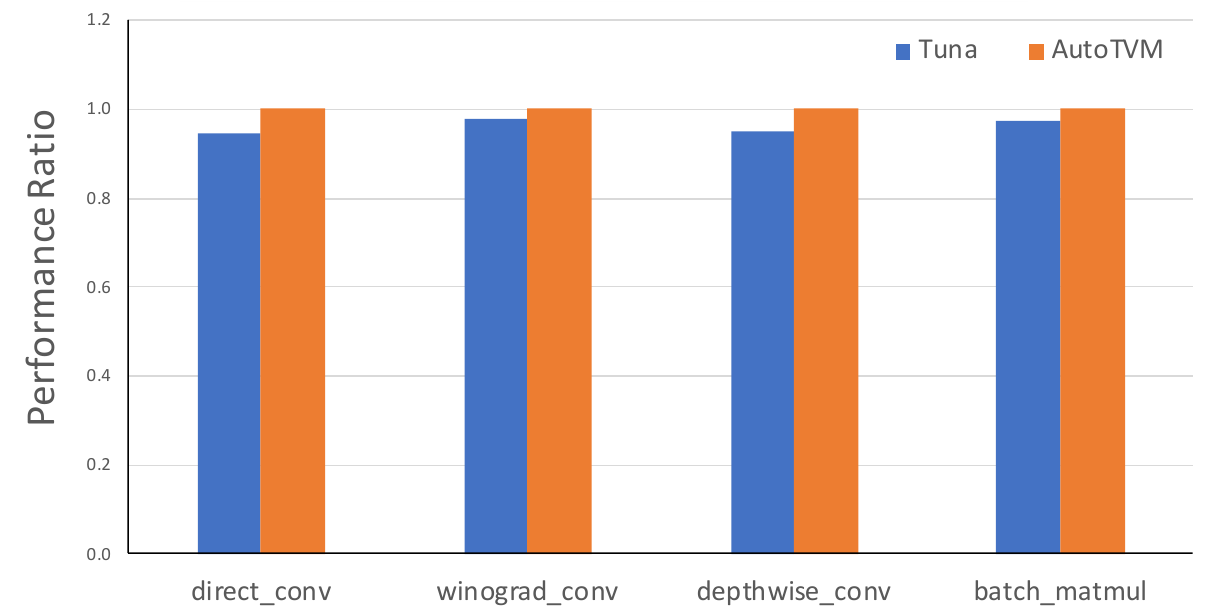}
		\caption{Top-50 Performance Ratio on AWS Graviton2 ARM CPU}
		\label{fig:single:top50-intel}
	\end{minipage}%
	\end{subfigure}
	\\
	\begin{subfigure}{0.47\textwidth}
	\begin{minipage}[t]{\textwidth}
		\centering
		\includegraphics[width=\linewidth]{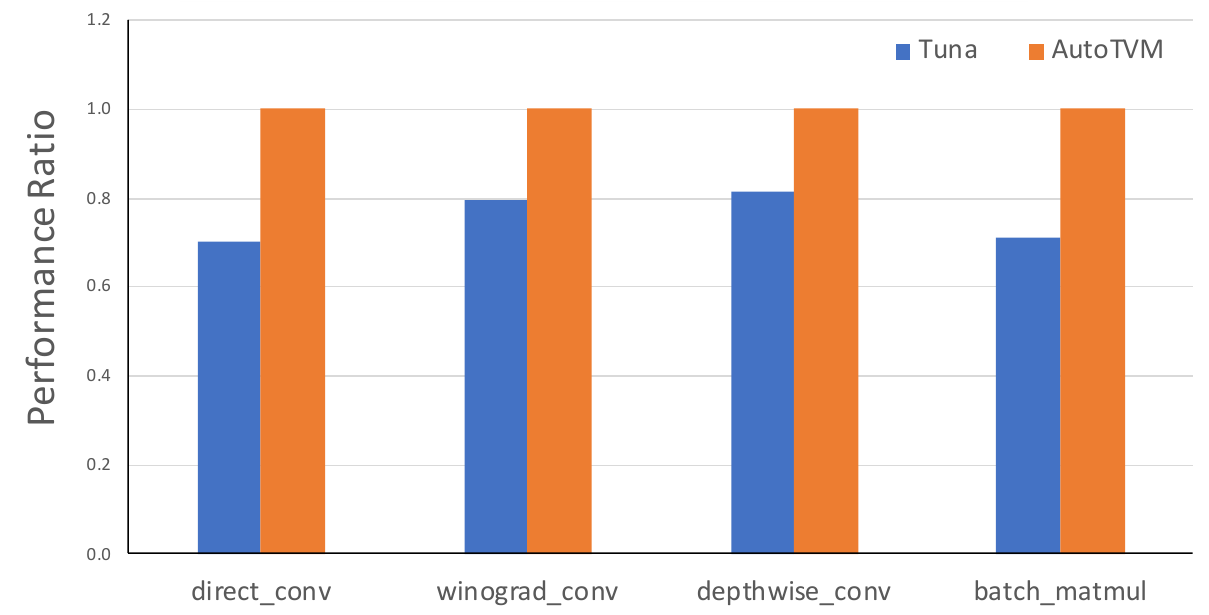}
		\caption{Top-50 Performance Ratio on AWS P3 V100 GPU}
		\label{fig:single:top50-intel}
	\end{minipage}%
	\end{subfigure}
	\caption{Top50 performance ratio for single operators from Tuna VS AutoTVM.}
	\label{fig:single50}
\end{figure}
In this paper, we need to search in a well-defined large search space to find the optimal configuration. We choose Evolution Strategies (ES) as the search algorithm and treat the search as an arbitrary black-box optimization problem. 

ES works by treating the model of interest as an arbitrary optimization problem. Given parameters of the model and the associated performance of that model on the task of interest, ES is able to optimize and train the model without any knowledge of the structure or architecture of the model itself. Specifically, at each iteration, random gaussian noise is added to the parameters to generate variations of the current model. Each variation is then tested for performance on the task of interest. Finally, a weighted sum is taken, based on performance, to generate an updated model. This becomes the current model for the next iteration. In standard ES, we treat the performance of the model within its environment as a black-box optimization problem. Now we extend that concept to treat ES itself as a black-box optimization problem with respect to the learning rate $\alpha$ and standard deviation of noise $\sigma$. 

\begin{algorithm}

{
  Get learning rate $\alpha$, noise standard deviation $\sigma$, initial policy parameters $\theta_0$\\
  \For {t = 0, 1, 2, ...}{
     Sample $\varepsilon_1, . . . \varepsilon_n \sim  \mathcal{N}(0, I)$ \\
     \For {i = 1, ..., n}{
       $F_i = F(\theta_t + \sigma \varepsilon_i)$
     }
     
     $\theta_{t+1} = \theta_t + \alpha\frac{1}{n\sigma}\sum_{i=1}^{n} F_i\varepsilon_i$ \\
  }
 }
  \caption{ Evolution Strategies }
  \label{algo:evostrategies}
\end{algorithm} 

The main steps about ES are given in Algorithm \ref{algo:evostrategies}, $F$ denotes objective function working with parameters $\theta$.
\section{Evaluation}
This section evaluates the performance of Tuna. The following hardware platforms are selected to cover the most widely used cases for deep learning inference: Intel Xeon Platinum 8124M CPU (Amazon EC2 C5.9xlarge), AWS Graviton2 (Amazon EC2 M6G.4xlarge, ARM 64-bit CPU), ARM Quad-core Cortex-A53 64-bit CPU (Acer aiSage), Nvidia Tesla V100 GPU (Amazon EC2 P3.2xlarge) and Nvidia Jetson AGX Xavier GPU (512-core Volta GPU with Tensor Cores).

We evaluate Tuna on two levels: single operator and entire neural network. We use AutoTVM as search-based tuning baseline for both levels of evaluation. We also compare Tuna with state-of-the-art deep learning framework solutions for the entire neural network evaluation.

We use TVM v0.7.0 release for all target devices. For Intel CPU targets, we chose TensorFlow 1.15.3 and Pytorch 1.6.0 as deep learning framework solutions. For AWS Graviton2 which has AARCH64 CPU, we use ARM Tool Solution docker container which provides AARCH64 optimized version of TensorFlow and Pytorch. For AWS P3 instance which has Nvidia V100 GPU, we use NGC docker container with tag 21.03-tf1-py3 and pytorch:21.03-py3 which optimizes TensorFlow and Pytorch for Nvidia GPU. For Nvidia Jetson AGX Xavier GPU, we choose TensorFlow 1.15.2 and Pytorch 1.6.0.

\subsection{Entire Network}
We first report Tuna performance on a set of deep learning models pre-trained with TensorFlow and PyTorch, two popular deep learning frameworks: TensorFlow SSD MobileNet v2, TensorFlow SSD Inception v2, PyTorch ResNet50 v1 and PyTorch Bert base uncased. Selected models cover the most popular deep learning applications: image classification, object detection and natural language processing. Three metrics are measures to demonstrate the advantages of Tuna:
\\ \\
\textbf{Compilation time.} Table~\ref{tbl:time} records the compilation time of both Tuna and AutoTVM to optimize neural networks. Experiments show that Tuna can speed up compilation process up to $339\times$ comparing to AutoTVM. 
\\ \\
\textbf{Compilation cost.} We measure the cost in dollar to optimize deep learning models on Amazon EC2 instances to demonstrate the cost reduction provided by Tuna. We multiply instance on-demand price with compilation time to get the cost to compile. We benchmark Tuna on Amazon EC2 C5.24xlarge which takes \$4.08 per hour. Table~\ref{tbl:cost} shows Tuna reduces compilation cost down to 1.1\% of the original cost comparing to AutoTVM.
\\ \\
\textbf{Inference latency.} Table~\ref{tbl:time} shows the model inference latency from different methods. The row of AutoTVM Partial shows the model inference latency compiled with AutoTVM in the same compilation time of Tuna. Results show that with the same compilation time, Tuna achieves up to $11\times$ performance of AutoTVM. We also compare Tuna latency with AutoTVM full tuning models. On average Tuna achieves 91.5\% performance comparing to best possible schedules generated by AutoTVM. This result shows that Tuna is able to achieve similar performance of full tuning with significant less time and cost. Finally we compare Tuna performance with deep learning framework solutions. Tuna achieves up to $17.3\times$ performance. Due to memory resource limitation on Acer aiSage device, directly running inference through deep learning framework is infeasible. We only compare Tuna with AutoTVM on this platform.

\subsection{Single Operator}
In this section we report the performance of a set of widely used compute-intensive deep learning operators. We use AutoTVM tuned operator performance as baseline. We reuse the optimization search space defined in AutoTVM to make apple to apple comparison. We benchmark $\mathit{conv2d}$, $\mathit{conv2d\_winograd}$, $\mathit{depthwise\_conv2d}$ and $\mathit{batch\_matrix\_multiplication}$ performance on ARM CPU and Nvidia GPU devices. We don't measure $\mathit{conv2d\_winograd}$ on Intel CPU device since AutoTVM doesn't define optimization space for this operator.

We define $\mathit{top-k}$ performance ratio as metric. Tuna generates top-k best programs and execution latency of all programs are sum up. Similarly we calculate the latency summation for top-k best programs generated with AutoTVM. We divide two AutoTVM latency value with Tuna to evaluate the effectiveness of Tuna to select decent optimizations from search space. Higher value approaching 1 indicates Tuna can accurately predict the real execution performance. Figure~\ref{fig:single10} and Figure~\ref{fig:single50} shows the benchmark result on Intel CPU, ARM CPU and Nvidia GPU. On average We got 0.869 for top 10 and 0.873 for top 50. Experimental data shows Tuna is able to achieve quite close performance comparing to full tuning method with AutoTVM.

\section{Related Works}
Performance modeling has been a challenging problem due to the diversities of hardware architectures. Over the years, researchers have proposed many analytical approaches. Here we summarize notable related works.
\\ \\
\textbf{CPU performance optimization:} 
Several state-of-the-art prior works has focused on analyzing the performance on CPU and generating efficient code.   They includes cache modeling tools \cite{bao2017analytical}\cite{gysi2019fast}, static tensor libraries\cite{low2016analytical}\cite{van2015blis}\cite{onednn}, model-driven tensor computation optimizers \cite{li2019analytical}\cite{li2021analytical}\cite{li2021efficient}, polyhedral compilers and polyhedral optimizations \cite{vasilache2018tensor}\cite{bondhugula2008practical}\cite{verdoolaege2013polyhedral}\cite{grosser2011polly}\cite{kong2013polyhedral}\cite{bastoul2004code}, and auto-tuning based compilers \cite{chen2018tvm}\cite{liu2019optimizing}\cite{zheng2020ansor}.  However, all of the state-of-the-art researches cannot be the off-the-shelf solution to our system for the following reasons. Cache miss analyzers are accurate on analyzing the cache misses but are not able to find the optimal loop schedule. Polyhedral compilers are usually separating the tile-size selection and loop transformation, which makes its performance not competitive to other frameworks.  Auto-tuning based compilers are tuning the loop schedule comprehensively for the whole network but results in very long tuning and compiling time.  Libraries are achieving high-performance on some important operators but their static optimization may not be optimal for all types of problem sizes, and they could block further graph-level optimizations such as fusion.  Model-driven tensor computation optimizers are efficient and comprehensive, but they are not yet fully automated and can only optimize specific computations such as tensor contractions and 2D convolutions.
\\ \\
\textbf{GPU performance modeling:}
Static analysis method has been used for GPU performance modeling. Arafa et al. \cite{arafa2019ppt} developed a GPU performance prediction framework named PPT-GPU, which utilizes a hybrid approach between analytical modeling and cycle-accurate modeling. They used features such as instruction statistics from the PTX code, instruction latencies, and L2 cache miss rate. Guerreiro et al. \cite{guerreiro2019gpu} proposed a GPU performance evaluation method based on recurrent neural networks and features extracted from PTX code. We need to emphasize the goal of Tuna is to sort and find the best kernels in a defined search space, while these approaches aim to accurately predict the performance of different kernels. One significant advantage of Tuna is that it leverages information from both low level assembly code and high level intermediate representations.
\\ \\
\textbf{Search-based auto-tuning:}
Search-based tuning approach has become an effective method to optimize deep learning programs. AutoTVM \cite{chen2018learning} relies on user-specified search space. Ansor \cite{zheng2020ansor} automatically generates search space for small sub-graphs. Both AutoTVM and Ansor create machine-learning based cost model and do training during exploration. Comparing to AutoTVM and Ansor, Tuna applies hardware feature related cost model which doesn't require any training and accurately predict relative performance. Halide auto-scheduler \cite{mullapudi2016automatically} generates schedules for the whole input program. All these search-based tuning methods require experimental execution on target device, while Tuna fully relies on static analysis.

\section{Conclusion}
In this paper, we proposed Tuna as an analytical approach to analyze and optimize deep learning programs on modern CPU and GPU hardware. The experiments show that we are able to achieve 99.21\% average throughput for three categories of deep learning models on various kinds of hardware platforms, with merely 1.65\% average compilation time comparing to search-based tuning approach. Tuna achieves 1.54x average performance comparing to state-of-the-art deep learning framework inference solutions. Similar methodology can be applied to ASIC hardware. Extending Tuna to other architecture is a future work.

\bibliographystyle{unsrt}
\balance
\bibliography{References.bib}

\balance
\end{document}